\documentclass[sigconf, nonatbib]{aamas}  

\usepackage{booktabs}

\usepackage{flushend}
\setcopyright{ifaamas}  
\acmDOI{doi}  
\acmISBN{}  
\acmConference[AAMAS'20]{Proc.\@ of the 19th International Conference on Autonomous Agents and Multiagent Systems (AAMAS 2020), B.~An, N.~Yorke-Smith, A.~El~Fallah~Seghrouchni, G.~Sukthankar (eds.)}{May 2020}{Auckland, New Zealand}  
\acmYear{2020}  
\copyrightyear{2020}  
\acmPrice{}  

\newcommand{\bE}{\mathbb{E}}
\newcommand{\bR}{\mathbb{R}}
\newcommand{\cA}{\mathcal{A}}
\newcommand{\cM}{\mathcal{M}}
\newcommand{\cS}{\mathcal{S}}

\newcommand{\cO}{\mathcal{O}}

\begin{document}

\title{Learning to Resolve Alliance Dilemmas in Many-Player Zero-Sum Games}  

\author{Edward Hughes}
\affiliation{
 \institution{DeepMind}
 \streetaddress{London, UK}
}
\email{edwardhughes@google.com}

\author{Thomas W. Anthony}
\affiliation{
 \institution{DeepMind}
 \streetaddress{London, UK}
}
\email{twa@google.com}

\author{Tom Eccles}
\affiliation{
 \institution{DeepMind}
 \streetaddress{London, UK}
}
\email{eccles@google.com}

\author{Joel Z. Leibo}
\affiliation{
 \institution{DeepMind}
 \streetaddress{London, UK}
}
\email{jzl@google.com}

\author{David Balduzzi}
\affiliation{
 \institution{DeepMind}
 \streetaddress{London, UK}
}
\email{dbalduzzi@google.com}

\author{Yoram Bachrach}
\affiliation{
 \institution{DeepMind}
 \streetaddress{London, UK}
}
\email{yorambac@google.com}

\begin{abstract} 
Zero-sum games have long guided artificial intelligence research, since they possess both a rich strategy space of best-responses and a clear evaluation metric. What's more, competition is a vital mechanism in many real-world multi-agent systems capable of generating intelligent innovations: Darwinian evolution, the market economy and the AlphaZero algorithm, to name a few. In two-player zero-sum games, the challenge is usually viewed as finding Nash equilibrium strategies, safeguarding against exploitation regardless of the opponent. While this captures the intricacies of chess or Go, it avoids the notion of cooperation with co-players, a hallmark of the major transitions leading from unicellular organisms to human civilization. Beyond two players, alliance formation often confers an advantage; however this requires trust, namely the promise of mutual cooperation in the face of incentives to defect. Successful play therefore requires adaptation to co-players rather than the pursuit of non-exploitability. Here we argue that a systematic study of many-player zero-sum games is a crucial element of artificial intelligence research. Using symmetric zero-sum matrix games, we demonstrate formally that alliance formation may be seen as a social dilemma, and empirically that na\"ive multi-agent reinforcement learning therefore fails to form alliances. We introduce a toy model of economic competition, and show how reinforcement learning may be augmented with a peer-to-peer contract mechanism to discover and enforce alliances. Finally, we generalize our agent model to incorporate temporally-extended contracts, presenting opportunities for further work.
\end{abstract}


\maketitle


\section{Introduction}\label{sec:background}

\textbf{Minimax foundations.} Zero-sum two-player games have long been a yardstick for progress in artificial intelligence (AI). Ever since the pioneering Minimax theorem \cite{v.Neumann1928, von_neumann_morgernstern_1948, shannon_1950}, the research community has striven for human-level play in grand challenge games with combinatorial complexity. Recent years have seen great progress in games with increasingly many states: both perfect information (e.g. backgammon \cite{Tesauro:1995:TDL:203330.203343}, checkers \cite{SCHAEFFER1992273}, chess, \cite{CAMPBELL200257}, Hex \cite{DBLP:journals/corr/AnthonyTB17} and Go \cite{Silver_2016}) and imperfect information (e.g. Poker \cite{DBLP:journals/corr/MoravcikSBLMBDW17} and Starcraft \cite{DBLP:journals/corr/abs-1708-04782}). 

The \textit{Minimax theorem} \cite{v.Neumann1928} states that every finite zero-sum two-player game has an optimal mixed strategy. Formally, if $x$ and $y$ are the sets of possible mixed strategies of the players and $A$ is the payoff matrix of the game, then $\max_x \min_y x^{\mathsf{T}} A y = \min_y \max_x x^{\mathsf{T}} A y = V$, with $V$ referred to as the value of the game. This property makes two-player zero-sum games inherently easier to analyze, as there exist optimal strategies which guarantee each player a certain value, no matter what their opponent does. 

In short, zero-sum two-player games have three appealing features:
\begin{enumerate}
    \item There is an unambiguous measure of algorithm performance, namely performance against a human expert.
    \item The size of the game tree gives an intuitive measure of complexity, generating a natural difficulty ordering for research.
    \item The minimax and Nash strategies coincide, so in principle there is no need for a player to adapt to another's strategy.
\end{enumerate}

\begin{figure}
    \centering
    \includegraphics[scale=0.3]{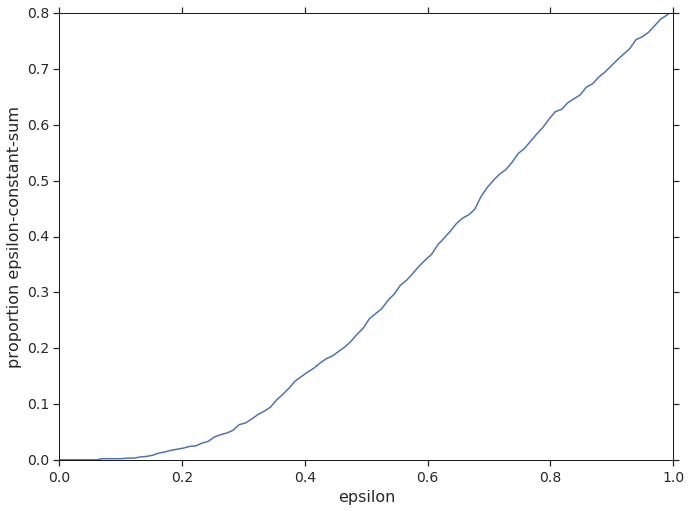}
    \caption{Two-player games were randomly generated with payoffs for each player drawn from $U(0, 1)$. A game is defined to be epsilon-constant-sum if the sums of both players' payoffs are mutually within $\ell_1$ distance epsilon of each other. $1000$ games were sampled and a histogram plotted.}
    \label{fig:constant_sum}
\end{figure}

{\bf Limitations of zero-sum two-player games}. The above properties of two-player zero-sum games makes them relatively easy to analyze mathematically, but most real-world interactions extend beyond direct conflict between two individuals. Zero-sum games are rare. Indeed, the chances of a random two-player two-action game being epsilon-constant-sum follow a triangular distribution, as shown empirically in Figure \ref{fig:constant_sum}. Importantly, this distribution has low density near $\epsilon = 0$: zero-sum games are even rarer than one might na\"ively expect. Thus, a research programme based around agent performance in two-player zero-sum games may fail to capture some important social abilities. More explicitly, property (3) is problematic: human intelligence is arguably predicated on our sociality \cite{reader_2002}, which exactly represents our ability to dynamically respond to a variety of co-players. Relatedly, natural environments are rarely purely adversarial; most environments are mixed-motive, where interactions between 
individuals comprise a combination of cooperation and competition. In other words, two-player constant-sum is a reasonable mathematical starting point for coarse-grained multi-agent interactions, but to understand the fine detail, we must move beyond this class of games. 

{\bf Approaches to multi-agent cooperation.} There are various ways to move beyond zero-sum two-player interactions. One profitable avenue is provided by zero-sum games between two teams of agents, as done in environments such as soccer~\cite{liu2018emergent} or Dota 2~\cite{OpenAI_dota}. Here, each team of agents faces an internal coordination problem. However, in principle this does not address the limitation of (3) at the level of the team's strategy. Moreover, since the teams are fixed, the individuals face no problem of team formation~\cite{gaston2005agent,sandholm1999coalition,bachrach2013optimal}, an important feature of socio-economic interactions, evinced by the substantial literature on cooperative game theory; see \cite{drechsel_2010} for a review. 

Alternatively, one might consider lifting the zero-sum restriction. Much recent work in multi-agent reinforcement learning (MARL) has focussed on social dilemmas with two or more players, e.g. \cite{DBLP:journals/corr/LeiboZLMG17, DBLP:journals/corr/PerolatLZBTG17, DBLP:journals/corr/abs-1803-08884, DBLP:journals/corr/abs-1811-05931, DBLP:journals/corr/abs-1903-08082, DBLP:journals/corr/LererP17, DBLP:journals/corr/abs-1709-04326}. These domains challenge learning agents to adapt and shape the learning of their co-players to achieve coordinated, cooperative outcomes without being exploited. In this sense, they directly address the limitations of (3). However, the general-sum case lacks the benefits conferred by properties (1) and (2). Indeed, there is no canonical metric of multi-agent performance, which makes comparisons between algorithms particularly difficult.

A third strand of work investigates scenarios in which cooperation is a given. Examples of such multi-agent learning problems include negotiation \cite{DBLP:journals/corr/abs-1804-03980}, coalition formation \cite{bachrach2019negotiating} and communication, both through a ``cheap talk'' channel \cite{DBLP:journals/corr/FoersterAFW16a,  DBLP:journals/corr/MordatchA17} and grounded in environment actions. These tasks certainly probe the abilities of agents to co-adapt, for they require intricate coordination. However, this is achieved at the expense of removing some of the adversarial dynamics. For example, in Hanabi any subset of agents are by definition incentivized to work together \cite{DBLP:journals/corr/abs-1902-00506}.

Finally, back in the two-player zero-sum setting, it is possible to optimize not for win-rate against an arbitrary opponent, but rather for maximum winning margin against some fixed background of exploitable opponents. For instance, in rock-paper-scissors against an opponent who always plays rock, you should always play paper, eschewing the mixed Nash. By definition one must adapt to one's co-players. Indeed, the identification of learning of robust best-responses has been extensively studied; see \cite{DBLP:journals/corr/abs-1906-01470, DBLP:journals/corr/abs-1711-00832, DBLP:journals/corr/abs-1906-01470, DBLP:journals/corr/abs-1901-08106} for a range of applications. Nevertheless, the necessity of choosing background opponents removes some of the advantages conferred by properties (1) and (2). 

\textbf{Our contributions}. The aims of this paper are threefold. Firstly to mathematically define the challenge of forming alliances. Secondly to demonstrate that state-of-the-art reinforcement learning, used to great effect in two-player zero-sum games, fails to form alliances. Thirdly, to provide a parsimonious and well-motivated mechanism for the formation of alliances by RL agents, namely contracts.

We focus on alliance formation in \textit{many-player} ($> 2$-player) zero-sum games as a fruitful and relatively understudied arena for AI research. More precisely, we examine symmetric zero-sum many-player games, and provide empirical results showing that alliance formation in such games often yields a social dilemma, thus requiring online adaptation between co-players. Viewing these games through the prism of reinforcement learning, we empirically show that standard independent reinforcement learning agents fail to learn to form alliances. We then propose a simple protocol that can be used to augment reinforcement learning agents, allowing them to learn peer-to-peer contracts, thus enabling them to make alliances. We study the impact of this protocol through a toy model of economic competition, demonstrating that contract-augment agents outperform the model-free baseline. Finally, we extend our model to incorporate enforcement by punishment, demonstrating the potential for our method to scale to spatially and temporally extended scenarios.

\section{Preliminaries}

\subsection{Related work}

We are far from the first to study many-player zero-sum games; see for instance \cite{kraus_1989, nijssen2013monte, brown2019superhuman, paquette2019press}. \citet{bonnet2018towards} have recently studied coalitions in Piglet, although they consider the coalition to be fixed from the beginning of the game. Our novel contributions come from studying alliance formation rigorously as a social dilemma, demonstrating the failure modes of MARL in alliance formation, and providing a mechanism for the learning of contracts. Despite much progress in this field, the problem of learning to form lasting, adaptable alliances without human data remains open. 

As inspiration, the cooperative game theory literature provides several methods for coordinating agents in competitive settings \cite{branzei2008models, chalkiadakis2011computational}. That line of work typically focuses on the question of how to share the joint gains of a team between the team members, in a fair way~\cite{bilbao2000generating,bachrach2010approximating,michalak2013efficient} or in a way that fosters the stability of the coalition~\cite{chalkiadakis2011computational,conitzer2006complexity,dunne2008cooperative,resnick2009cost}. Such work has also been used as a foundation for building agents that cooperate and negotiate with humans~\cite{kraus1997negotiation,jennings2001automated,mash2017form,rosenfeld2018predicting}. In particular, researchers have proposed algorithms for robust team formation~\cite{shehory1998methods, gaston2005agent, klusch2002dynamic, zlotkin1994coalition} and sharing resources \cite{choi2009consensus, rosenschein1994rules, shamma2007cooperative}. Our work builds on these strong foundations, being the first to apply MARL in this setting, and demonstrating the efficacy of a concrete protocol that allows MARL agents to form alliances. 

Our MARL approach augments agents with the ability to negotiate and form contracts regarding future actions. Many protocols have been suggested for multi-agent negotiation, as discussed in various surveys on the topic \cite{shamma2007cooperative, kraus1997negotiation, rosenschein1994rules, kraus2001strategic}. Some propose agent interaction mechanisms \cite{smith1980contract, kuwabara1995agentalk, sandholm1995issues, sandholm1996advantages, fornara2002operational}, whereas others aim to characterize the stability of fair outcomes for agents \cite{tsvetovat2000customer, klusch2002dynamic, conitzer2004computing, ieong2005marginal}. However, the focus here lies on solution concepts, rather than the online learning of policies for both acting in the world and developing social ties with others.

As MARL has become more ubiquitous and scalable, so has interest in dynamic team formation increased. In the two-player setting, the Coco-$Q$ algorithm achieves social welfare maximizing outcomes on a range of gridworld games when reinforcement learning agents are augmented with the ability to sign binding contracts and make side payments \cite{Sodomka}. For many-agent games, there has been much progress on achieving adhoc teamwork \cite{Stone2010AdHA} in fully cooperative tasks, including by modelling other agents \cite{Barrett2012LearningTM}, invoking a centralized manager \cite{shu2018mrl} or learning from past teammates \cite{BARRETT2017132}. To our knowledge, no previous literature has studied how alliances may be formed by learning agents equipped with a contract channel in many-agent zero-sum games. 

As this paper was in review, two complementary works appeared. \cite{jonge} equips agents with the ability to negotiate and form contracts as part of a search algorithm, aligned with yet distinct from our reinforcement learning approach. \cite{alex2019colosseumrl} defines a framework for reinforcement learning in $N$-player games, which could be used to scale up our results, but does not explicitly tackle the issue of alliance formation.

\subsection{Multi-agent reinforcement learning}

We consider multi-agent reinforcement learning (MARL) in Markov games \cite{shapley1953stochastic, Littman94markovgames}. In each game state, agents take actions based on a (possibly partial) observation of the state space and receive an individual reward. Agents must learn an appropriate behavior policy through experience while interacting with one another. We formalize this as follows.

Let $\cM$ be an $n$-player partially observable Markov game defined on a finite set of states $\cS$. The observation function $O: \cS \times \{1,\dots,n\} \rightarrow \bR^d$ specifies each player's $d$-dimensional view on the state space. From each state, each player may take actions from each of the sets $\cA^1,\dots,\cA^n$ respectively. As a result of the joint action $(a^1,\dots,a^n) \in \cA^1\times\dots\times\cA^n$ the state changes, following the stochastic transition function $T: \cS \times \cA^1 \times \cdots \times \cA^n \rightarrow \Delta(\cS)$, where $\Delta(\cS)$ denotes the set of discrete probability distributions over $\cS$. We denote the observation space of player $i$ by $\cO^i = \{ o^i~|~s \in \cS, o^i = O(s,i)\}$. Each player receives an individual reward denoted by $r^i: \cS \times \cA^1 \times \cdots \times \cA^N \rightarrow \bR$ for player $i$.  We denote the joint actions, observations and policies for all players by $\vec{a}$, $\vec{o}$ and $\vec{\pi}$ respectively.

In MARL, each agent learns, independently through its own experience of the environment, a behavior policy $\pi^i : \cO^i \rightarrow \Delta(\cA^i)$, which we denote $\pi(a^i|o^i)$. Each agent's goal is to maximize a long term $\gamma$-discounted payoff,\footnote{For all of our experiments, $\gamma = 0.99$.} namely
\begin{equation}
V_{\vec{\pi}}^i(s_0) = \bE \left( \sum \limits_{t=0}^{\infty} \gamma^t r^i(s_t, \vec{a}_t)  \right) \, ,
\end{equation}
where $\vec{a}_t$ is a sampled according to the distribution $\vec{\pi}_t$ and $s_{t+1}$ is sampled according to the distribution $T(s_t, \vec{a}_t)$. In our setup, each agent comprises a feedforward neural network and a recurrent module, with individual observations as input, and policy logits and value estimate as output. We train our agents with a policy gradient method known as advantage actor-critic (A2C) \cite{Sutton1998}, scaled up using the IMPALA framework \cite{DBLP:journals/corr/abs-1802-01561}. Details of neural network configuration and learning hyperparameters are provided alongside each experiment to aid reproducibility. 

\section{Alliance Dilemmas}\label{sec:alliance-dilemmas}

In this section we formally introduce the notion of an alliance dilemma, a definition which captures the additional complexity present in $n$-player zero-sum games when $n > 2$.

\textbf{Intuition}. The simplest class of multi-player zero-sum games are the symmetric $2$-action, $3$-player repeated matrix games. Particularly interesting games in this class are those which reduce to two-player social dilemmas if one of the participants adheres to a fixed policy. Here, one agent can make the team formation problem hard for the other two, and may be able to unreasonably generate winnings, since independent reinforcement learning algorithms in social dilemmas converge to defecting strategies. In other words, a appropriately configured ``stubborn'' agent can force the two dynamic agents into a social dilemma, whence MARL finds the worst collective outcome for them: the Nash.\footnote{If each agent is running a no-regret algorithm then the average of their strategies converges to a coarse-correlated equilibrium \cite{Gordon:2008:NLC:1390156.1390202}. Empirically, there is much evidence in the literature to suggest that reinforcement learning does reliably find the Nash in social dilemma settings, but this is not theoretically guaranteed.}

\textbf{Social dilemmas}.
A two-player matrix game,
\vspace{4mm}
\begin{center}
\begin{tabular}{ | c | c | } 
\hline
R, R & S, T \\ 
\hline
T, S & P, P \\ 
\hline
\end{tabular}
\end{center}
\vspace{4mm}
is called a \textit{social dilemma} if and only if \cite{chammah1965prisoner, Macy7229}:
\begin{enumerate} 
    \item $R > P$: mutual cooperation is preferred to mutual defection.
    \item $R > S$: mutual cooperation is preferred to being exploited by a defector.
    \item either $T > R$: exploiting a cooperator is preferred over mutual cooperation (\textit{greed}), or $P > S$: mutual defection is preferred over being exploited by a defector (\textit{fear}).
\end{enumerate}
We say that the social dilemma is \textit{strict} iff in addition
\begin{enumerate}
    \setcounter{enumi}{3}
    \item $2R > T + S$: mutual cooperation is preferred over an equal mix of defection and cooperation.
\end{enumerate}
There is some debate as to the importance of this final condition: for example, several papers in the experimental literature violate it \cite{beckenkamp}. We shall demonstrate that pathological behavior arises in gradient-based learning even in non-strict social dilemmas. Thus from the perspective of alliances, the first three conditions should be considered to be the important ones. 

\textbf{Many-player zero-sum games}.
An $n$-player $2$-action matrix games is called \textit{zero-sum} if and only if for each vector of simultaneous payoffs $(U_1, \dots U_n)$, the following holds: $\sum_{i=1}^n U_i = 0$. In other words, a gain in utility for any one player must be exactly balanced by a loss in utility for the rest. A well-known corollary of this is that every outcome is Pareto efficient. However, from the perspective a $k$-player alliance ($k < n$) within the zero-sum game, Pareto improvements are possible. Note that many-player zero-sum is strictly more general than the class of pairwise zero-sum games recently studied in the context of learning \cite{DBLP:journals/corr/abs-1901-08106}. Importantly, we make no restriction on the payoffs for interactions between strict subsets of players.

\textbf{Alliance dilemmas}.
An \textit{alliance dilemma} in an $n$-player zero-sum $2$-action matrix game is a social dilemma which arises for a $2$-player subset, on the basis of the current policies of their co-players. Mathematically, denote the action space for player $i$ by $\{a_i^0, a_i^1\}$ with $1 \leq i \leq n$. An alliance dilemma is a social dilemma for players $i$ and $j$ which arises when the game is reduced according to some fixed policies for the rest of the players $\pi_k(a_k^0, a_k^1)$ for $k \neq i, j$. 

It is not immediately clear whether such situations are ubiquitous or rare. In this section we shall show that at least in the atomic setting described above, alliance dilemmas arise often in randomly generated games. Moreover, we shall justify the informal statement that gradient-based learning fails to achieve reasonable outcomes in the presence of alliance dilemmas, by considering the learning dynamics in two easily interpretable examples.

\subsection{Counting alliance dilemmas}

How often can a stubborn agent in symmetric three-player two-action matrix games generate a social dilemmas for their two opponents? To answer this question we randomly sample $1000$ such games, parameterized by $0 \leq p \leq 1$ and $0 \leq q \leq 1$ as follows:

\begin{center}
\begin{tabular}{|c|c|}
    \hline
     Actions & Payoffs \\
     \hline
     $(0, 0, 0)$ & $(\frac{1}{3}, \frac{1}{3}, \frac{1}{3})$ \\
     \hline
     $(0, 0, 1)$ & $(\frac{1 - p}{2}, \frac{1 - p}{2}, p)$ \\
     \hline
     $(0, 1, 0)$ & $(\frac{1 - p}{2}, p, \frac{1 - p}{2})$ \\
     \hline
     $(0, 1, 1)$ & $(q, \frac{1 - q}{2}, \frac{1 - q}{2})$ \\
     \hline
     $(1, 0, 0)$ & $(p, \frac{1 - p}{2}, \frac{1 - p}{2})$ \\
     \hline
     $(1, 0, 1)$ & $(\frac{1 - q}{2}, q, \frac{1 - q}{2})$ \\
     \hline
     $(1, 1, 0)$ & $(\frac{1 - q}{2}, \frac{1 - q}{2}, q)$ \\
     \hline
     $(1, 1, 1)$ & $(\frac{1}{3}, \frac{1}{3}, \frac{1}{3})$ \\
     \hline
\end{tabular}
\end{center}

\begin{figure}
    \centering
    \includegraphics[scale=0.4]{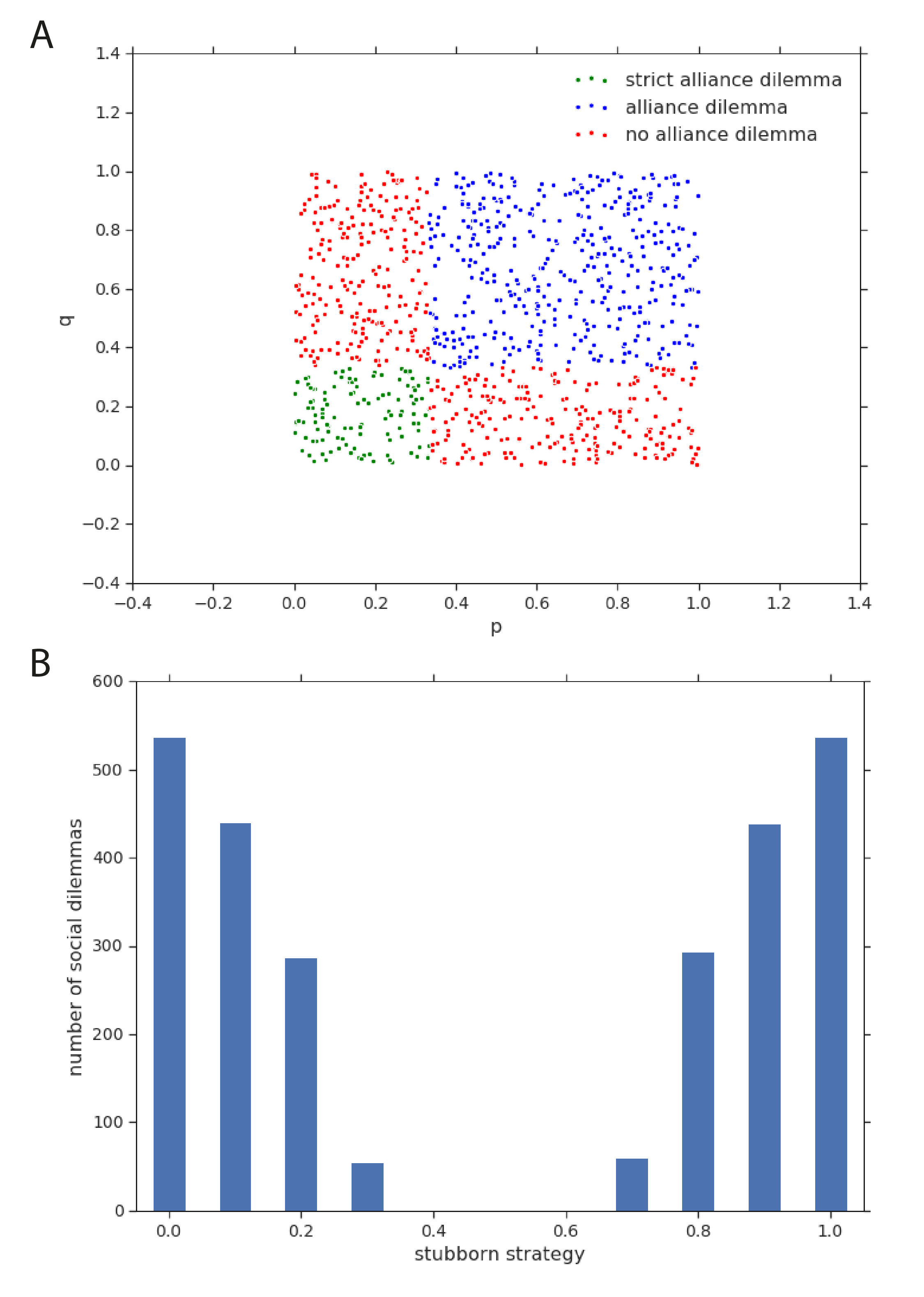}
    \caption{The presence of absence of alliance dilemmas in $1000$ simulated three-player zero-sum matrix games. (A) Alliance dilemmas arise for characteristic values of the parameters $p$ and $q$ of the three-player zero-sum game. (B) Alliance dilemmas become more common as the policy of the stubborn agent becomes more deterministic.}
    \label{fig:simulation}
\end{figure}

For each game we consider $11$ possibilities for the stubborn player's policy, given by discretizing the probability of taking action $0$ with step size $0.1$. For each of these we compute the resulting matrix game for the remaining players and check the social dilemma conditions. If any of these yield a social dilemma, then the original game is an alliance dilemma, by definition. Overall we find that 54\% of games contain an alliance dilemma, with 12\% of these being strict alliance dilemmas. Figure \ref{fig:simulation}(A) depicts the distribution of these alliance dilemmas in $p$-$q$ space. The striking structure is easily seen to be the effect of the linear inequalities in the definition of a social dilemma. Figure \ref{fig:simulation}(B) demonstrates that alliance dilemmas arise for stubborn policies that are more deterministic. This stands to reason: the stubborn player must leach enough of the zero-sum nature of the game to leave a scenario that calls for cooperation rather than pure competition. 

\subsection{The failure of gradient-based learning}

There are two values of $p$ and $q$ which result in easily playable games. When $p = 1$ and $q =0$ we obtain the \textit{Odd One Out} game. Here, you win outright if and only if your action is distinct from that of the other two players. Similarly, the combination $p = 0$ and $q = 1$ defines the \textit{Matching} game. Here, you lose outright if and only if your action is distinct from that of the other two players. It is easy to verify that both of these games contain an alliance dilemma. Odd One Out has a non-strict greed-type dilemma, while Matching has a strict fear-type dilemma. This classification is intuitively sensible: the dilemma in Odd One Out is generated by the possibility of realizing gains, while the dilemma in Matching arises because of the desire to avoid losses.

\begin{figure}
    \centering
    \includegraphics[scale=0.5]{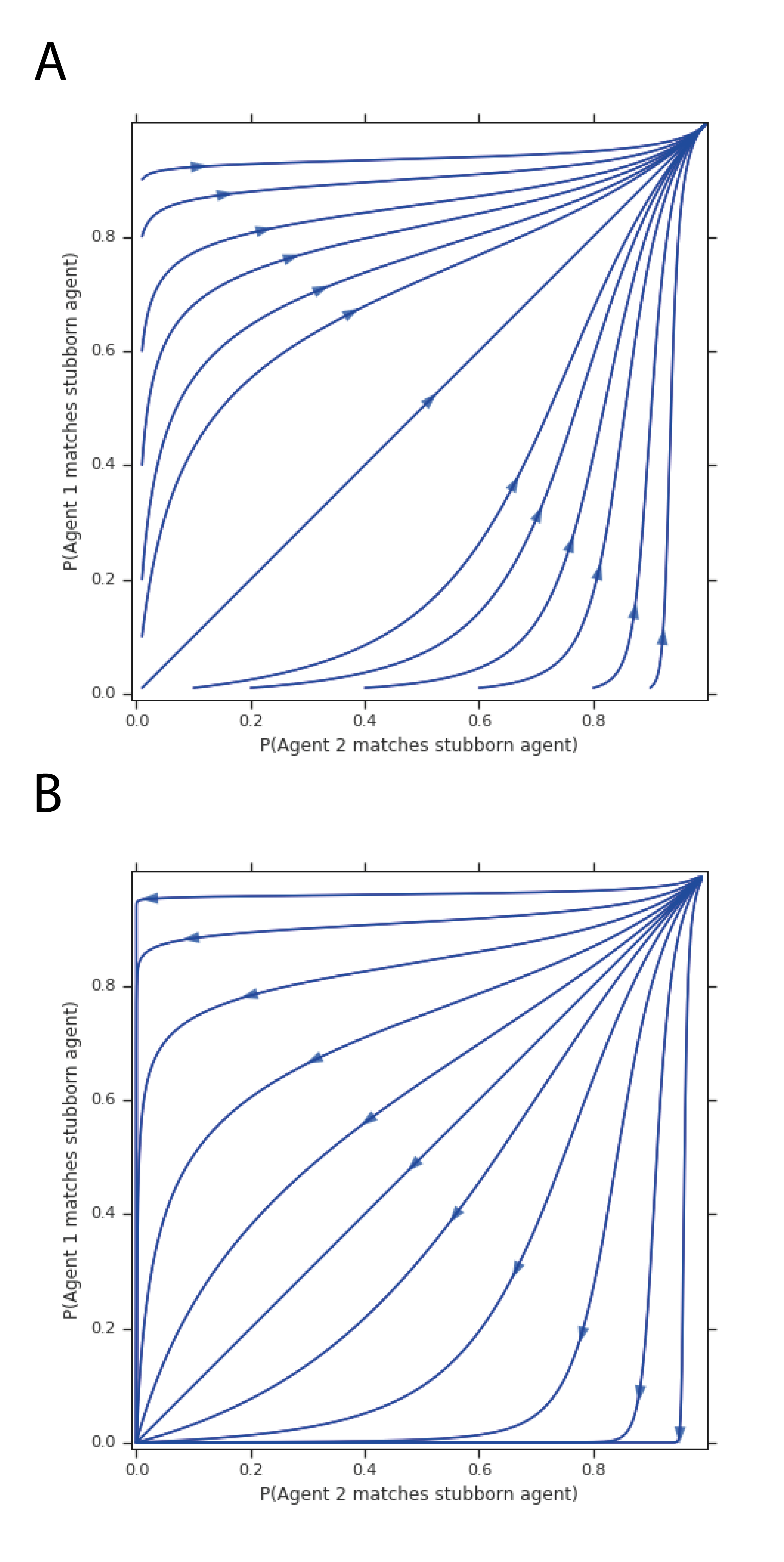}
    \caption{The learning dynamics of two example games. The lines shown are the trajectories of agents under softmax policy gradients with the same learning rate. In Matching (A), both agents learn to match the stubborn agent. In Odd One Out (B), neither agent learns to match the stubborn agent.}
    \label{fig:rl-odd-one-out}
\end{figure}

Despite the simplicity of these games, gradient-based learning fails to find alliances when pitched against a stubborn agent with a deterministic policy (see Figure \ref{fig:rl-odd-one-out}). More precisely, in Matching the optimal solution for agents in an alliance is to match actions, taking the opposite action from the stubborn agent to gain reward $\frac{1}{2}$ each. However, policies initialised anywhere other than these optimal policies converge to both taking the same action as the stubborn agent, each getting reward $\frac{1}{3}$. In Odd One Out, the optimal symmetric alliance solution is to each match the stubborn agent $\frac{3}{4}$ of the time, for an average reward of $\frac{3}{8}$.\footnote{If non-stubborn agents both match the stubborn agent with probability $p$, their expected joint return is $\frac{2}{3}p^2 + 2 p(1-p)$.} The learning dynamics of the system tend towards a fixed point in which both agents never match the stubborn agent, gaining no reward at all. In this game, we also see that the system is highly sensitive to the initial conditions or learning rates of the two agents; small differences in the starting point lead to very different trajectories. We provide a mathematical derivation in Appendix \ref{appendix:fail} which illuminates our empirical findings. 

Despite the fact that Matching yields a strict social dilemma while Odd One Out gives a non-strict social dilemma, the learning dynamics in both cases fail to find the optimal alliances. As we anticipated above, this suggests that condition (4) in the definition of social dilemmas is of limited relevance for alliance dilemmas.

\subsection{Gifting: an alliance dilemma}

In the previous section we saw that alliance dilemmas are a ubiquitous feature of simultaneous-move games, at least in simple examples. We now demonstrate that the same problems arise in a more complex environment. Since our definition of alliance dilemma relies on considering the policy of a given agent as fixed at each point in time, we shift our focus to sequential-move games, where such analysis is more natural, by dint of only one player moving at a time. 

\begin{figure}
    \centering
    \includegraphics[scale=0.25]{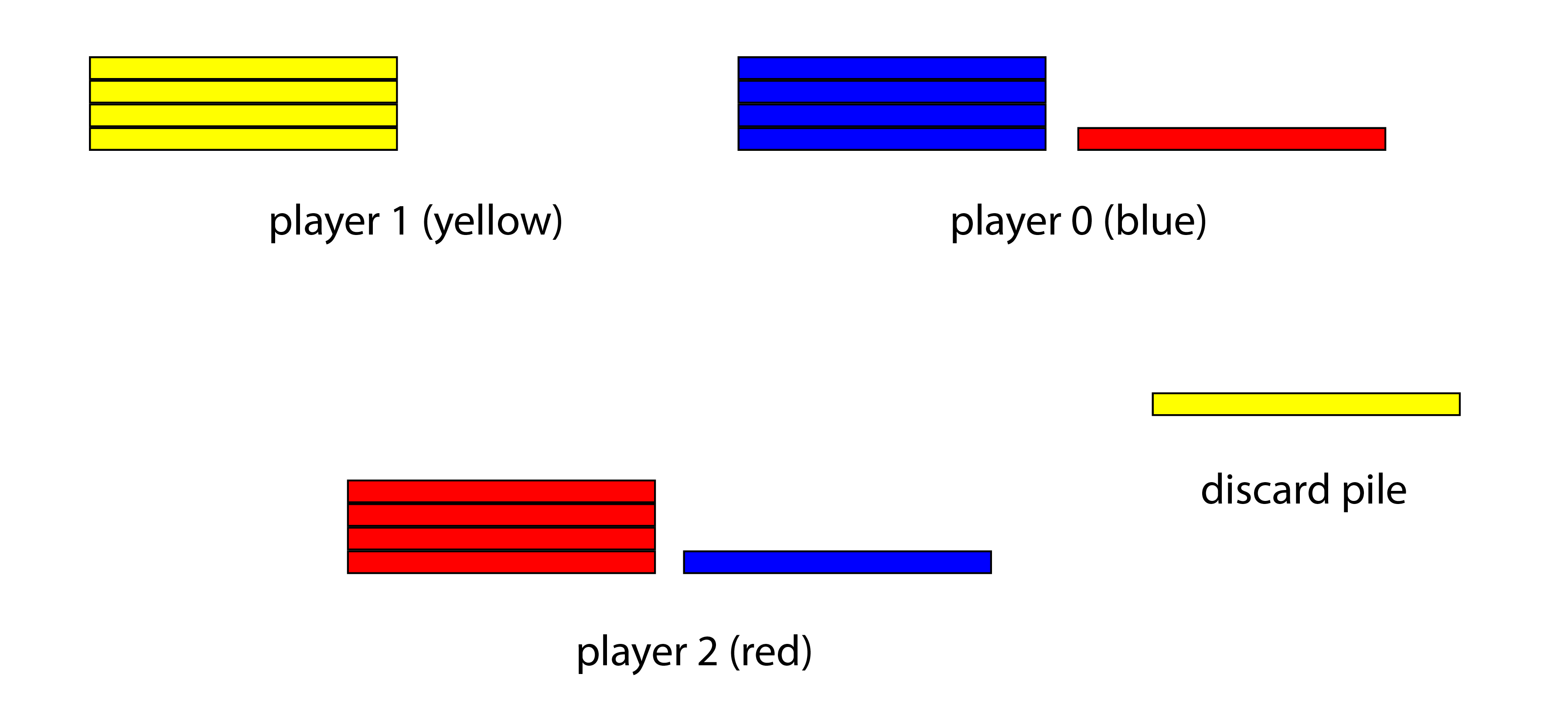}
    \caption{The Gifting game. Each player has taken $1$ turn. Player $0$ gifted a blue chip to player $2$; player $1$ discarded a yellow chip; player $2$ gifted a red chip to player $0$.}
    \label{fig:gifting}
\end{figure}

In the \textit{Gifting} game (see Figure \ref{fig:gifting}), each player starts with a pile of $m$ chips of their own colour (for our experiments, $m = 5$). On a player's turn they must take a chip of their own colour and either gift it to another player or discard it from the game. The game ends when no player has any chips of their own color left; that is to say, after $n \times m$ turns. The winner is the player with the most chips (of any colour), with $k$-way draws possible if $k$ players have the same number of chips. The winners share a payoff of $1$ equally, and all other players receive a payoff of $0$. Hence Gifting is a constant-sum game, which is strategically equivalent to a zero-sum game, assuming no externalities. Moreover, it admits an interpretation as a toy model of economic competition based on the distribution of scarce goods, as we shall see shortly.

That ``everyone always discards'' is a subgame perfect Nash equilibrium is true by inspection; i.e. no player has any incentive to deviate from this policy in any subgame, for doing so would merely advantage another player, to their own cost. On the other hand, if two players can arrange to exchange chips with each other then they will achieve a two-way rather than a three-way draw. This is precisely an alliance dilemma: two players can achieve a better outcome for the alliance should they trust each other, yet each can gain by persuading the other to gift a chip, then reneging on the deal. Accordingly, one might expect that MARL fails to converge to policies that demonstrate such trading behavior.

\begin{figure}
    \centering
    \includegraphics[scale=0.38]{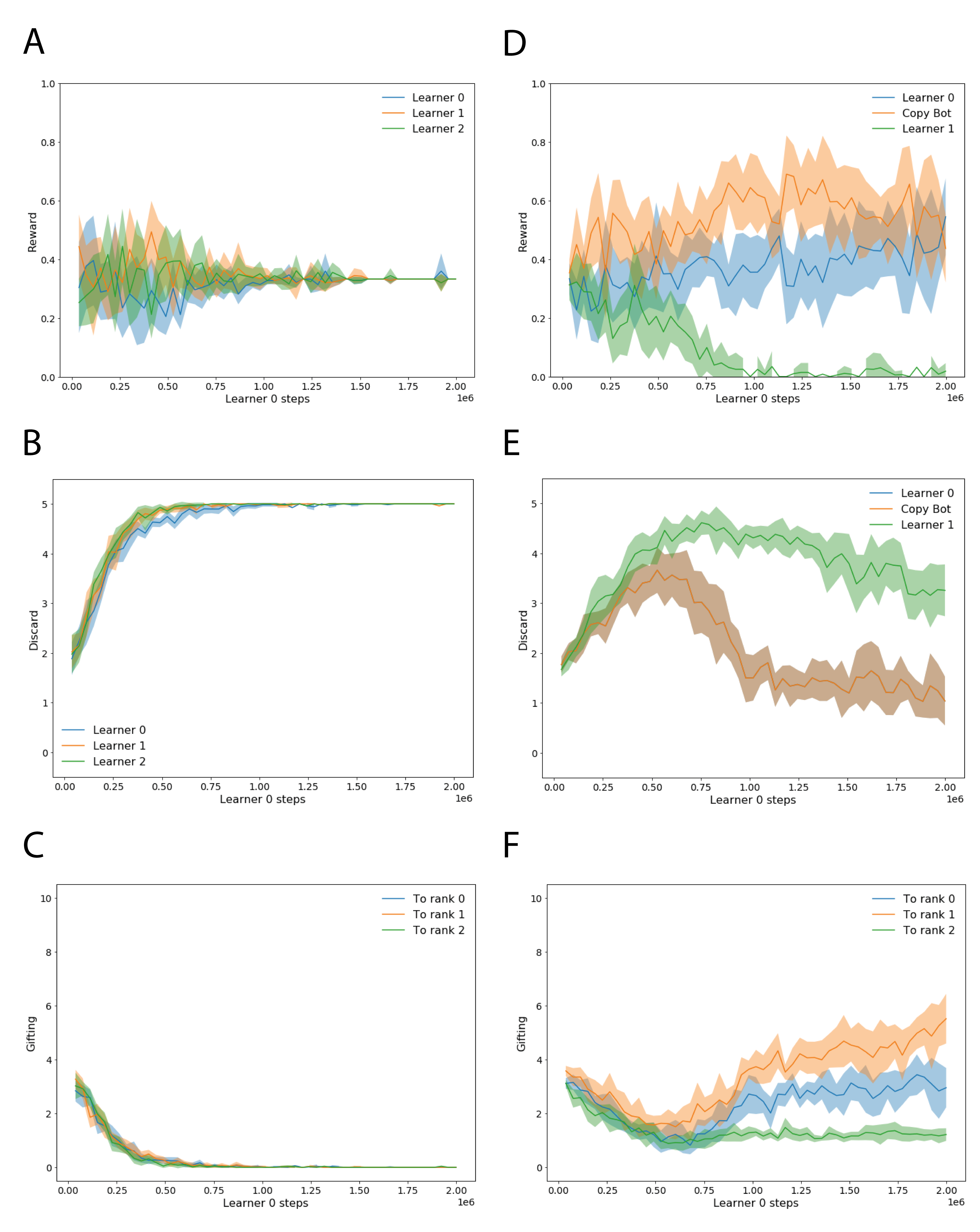}
    \caption{Learning curves for MARL on the Gifting environment. (A--C) Three independent learners quickly converge to the subgame perfect Nash of never gifting, resulting in a three-way draw. (D--F) One of the learners is replaced by a copy bot which always mimics the actions of the first learner. The first learner now learns to gift, since this will surely be reciprocated. The result is a two-way draw. Dependent variables are as follows. (A) Reward. (B) Discard. (C) Gifting. (D) Reward. (E) Discard. (F) Gifting.}
    \label{fig:rl-gifting}
\end{figure}

\textbf{Results}.\footnote{In each episode, agents are randomly assigned to a seat $1$, $2$ or $3$, so must generalize over the order of play. Each agent's neural network comprises an MLP with two layers of $128$ hidden units, followed by an LSTM with $128$ hidden units. The policy and value heads are linear layers on top of the LSTM. We train with backpropagation-through-time, using an unroll length equal to the length of the episode. Optimization is carried out using the RMSProp optimizer, with decay $0.99$, momentum $0$, epsilon $0.001$ and learning rate $0.000763$. The entropy cost for the policy is $0.001443$. We perform training runs initialized with $10$ different random seeds and plot the average with 95\% confidence intervals.} This expectation is borne out by the results in Figure \ref{fig:rl-gifting}(A--C). Agents start out with different amounts of discarding behavior based on random initialization, but this rapidly increases. Accordingly, gifting behavior to agents decreases to zero during learning. The result is a three-way draw, despite that fact that two agents that agreed to exchange could do better than this. To demonstrate this final point, we replace the second player with a bot which reciprocates the gifting behavior of player $0$; see Figure \ref{fig:rl-gifting}(D--F). Under these circumstances, player $0$ learns to gift to player $1$, leading to a two-way draw. Player $2$ initially learns to discard, but soon this behavior confers no additional reward, at which point the entropy regularizer leads to random fluctuation in the actions chosen. 

We conclude that reinforcement learning is able to adapt, but only if an institution supporting cooperative behavior exists. MARL cannot create the kind of reciprocal behavior necessary for alliances ex nihilo. Inspired by the economics literature, we now propose a mechanism which solves this problem: learning to sign peer-to-peer contracts.   

\section{Contracts}

\subsection{Binding contracts}

The origin of alliance dilemmas is the greed or fear motivation that drives MARL towards the Pareto-inefficient Nash equilibrium for any given $2$-player subset.\footnote{See the definition of social dilemma in Section \ref{sec:alliance-dilemmas} for a reminder of these concepts.}  Humans are adept at overcoming this problem, largely on the basis of mutual trust \cite{van2017trust}. Much previous work has considered how inductive biases, learning rules and training schemes might generate trust in social dilemmas. Viewing the problem explicitly as an alliance dilemma yields a new perspective: namely, what economic mechanisms enable self-interested agents to form teams? A clear solution comes from contract theory \cite{Martimort2017}. Binding agreements that cover all eventualities trivialize trust, thus resolving the alliance dilemma.

\textbf{Contract mechanism}. We propose a mechanism for incorporating peer-to-peer pairwise complete contracts into MARL. The core environment is augmented with a contract channel. On each timestep, each player $i$ must submit a contract offer, which comprises a choice of partner $j$, a suggested action for that partner $j$, and an action which $i$ promises to take, or no contract. If two players offer contracts which are identical, then these become binding; that is to say, the environment enforces that the promised actions are taken, by providing a mask for the logits of the relevant agents. At each step, agents receive an observation of the contracts which were offered on the last timestep, encoded as a one-hot representation. 

\textbf{Contract-aware MARL}. To learn in this contract-augmented environment, we employ a neural network with two policy heads, one for the core environment and another for the contract channel. This network receives the core environment state and the previous timestep contracts as input. Both heads are trained with the A2C algorithm based on rewards from the core environment, similarly to the RIAL algorithm \cite{DBLP:journals/corr/FoersterAFW16a}. Therefore agents must learn simultaneously how to behave in the environment and how to leverage binding agreements to coordinate better with peers. 

\begin{figure}
    \centering
    \includegraphics[scale=0.38]{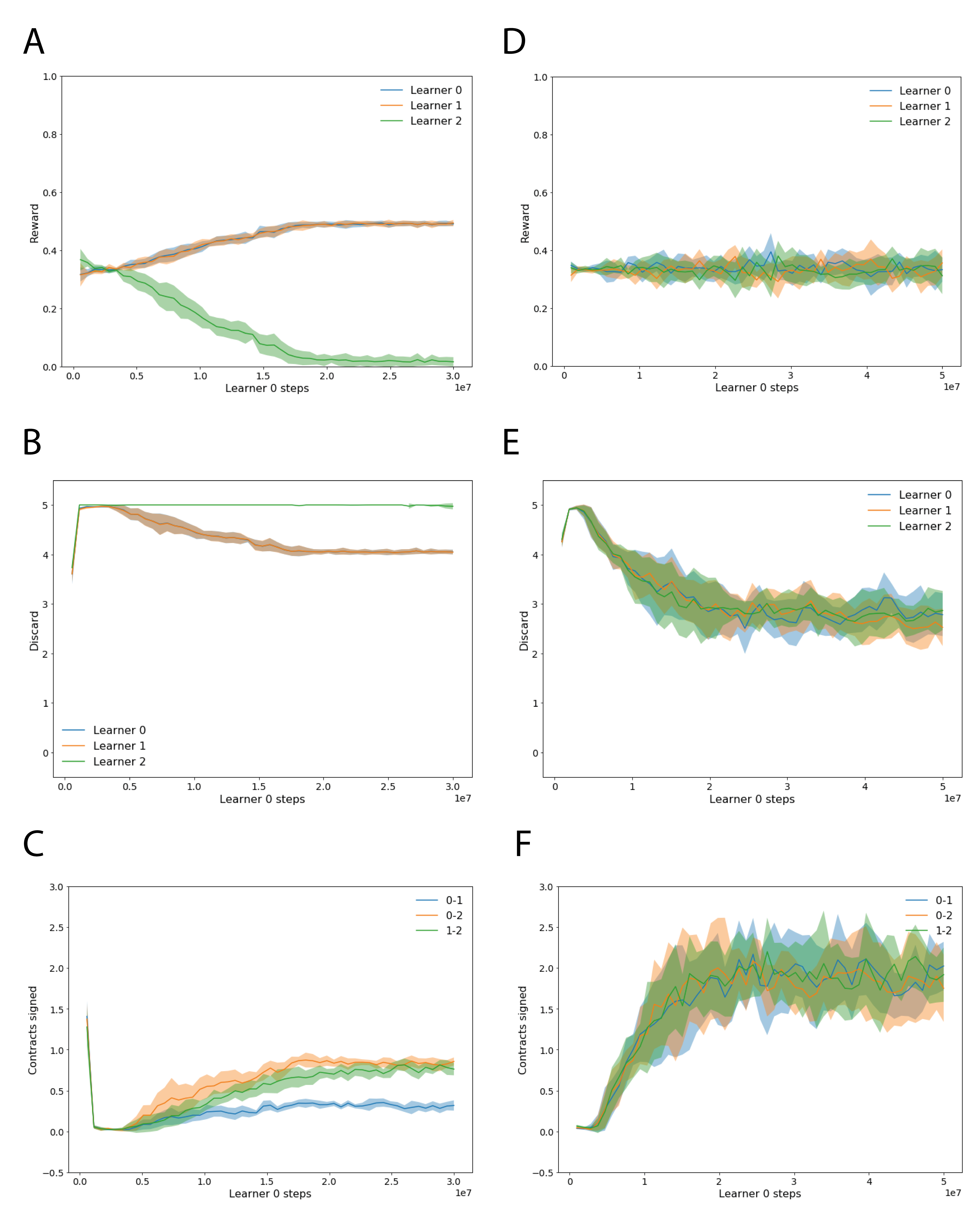}
    \caption{Performance of contract-augmented agents on Gifting. Data comes from evaluation episodes without a trembling hand policy. (A--C) Learners $0$ and $1$ are endowed with the ability to make contracts, but leaner $2$ is not. The contract augment agents use this ability to consistently achieve a $2$-way draw. (D--F) All learners are endowed with the ability to make contracts. (D) Rewards across training become more chaotic, since agents are vying with each other to sign the contracts to secure a $2$-way draw (D). As agents compete to sign contracts guaranteeing them more gifts, the number of discards decreases (E) and the number of contracts signed increases (F). Dependent variables are as follows. (A) Reward. (B) Discard. (C) Contracts signed. (D) Reward. (E) Discard. (F) Contracts signed.}
    \label{fig:gifting-contracts}
\end{figure}

\begin{figure}
    \centering
    \includegraphics[scale=0.5]{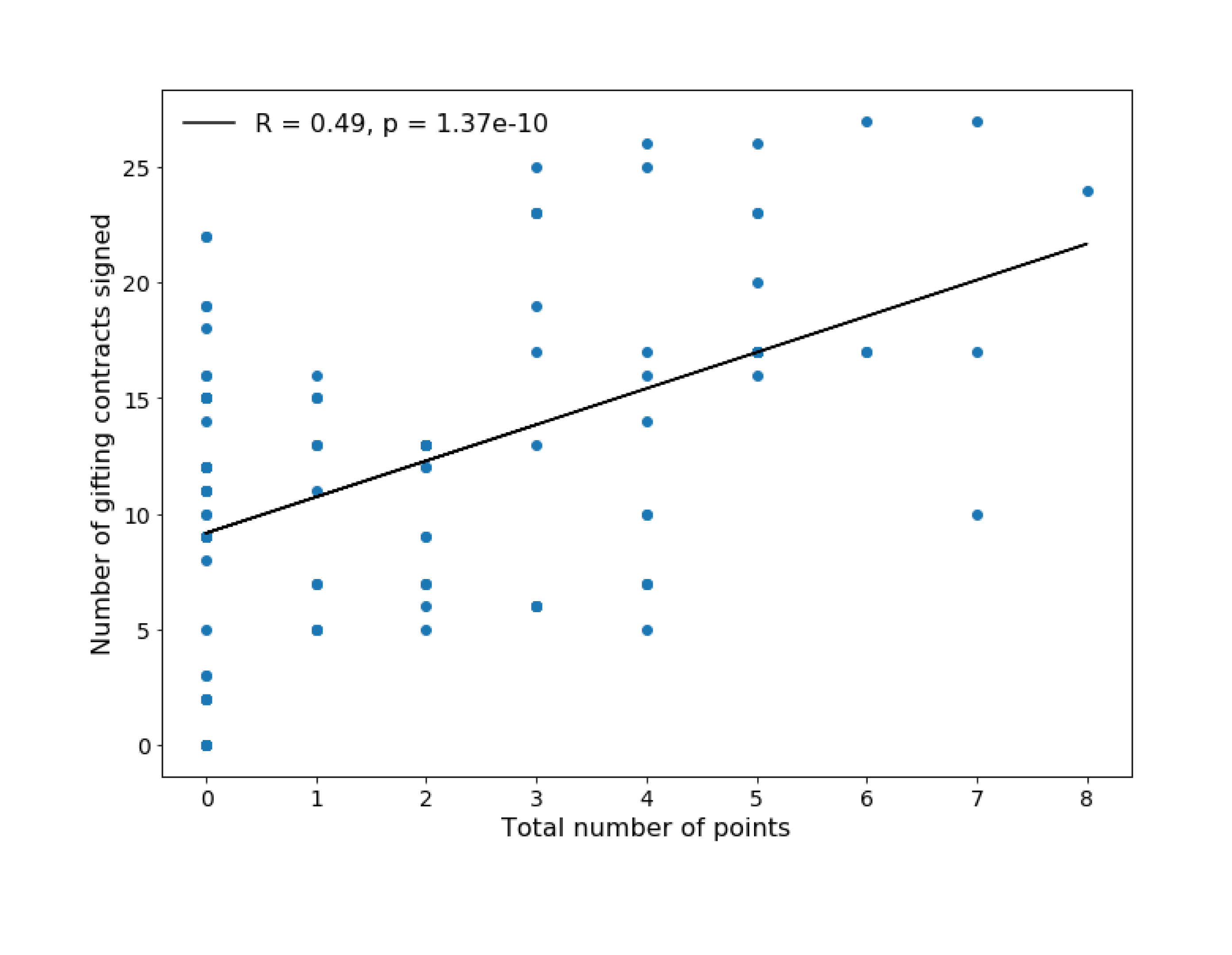}
    \caption{$50$ games are rolled out in evaluation (after training). For each game, we record the total number of chips accrued (points) and contracts signed by each player. We present a scatter plot of all $150$ such values. Linear regression shows a significant positive correlation between the signing of gifting contracts and points accrued during the game.}
    \label{fig:regression}
\end{figure}

\textbf{Results}.\footnote{The neural network is identical to the baseline experiments, except for the addition of a linear contract head. The optimizer minimizes the combined loss $L_{\textrm{RL}} + \alpha L_{\textrm{contract}}$ where $\alpha = 1.801635$. We include an entropy regularizer for the contract policy, with entropy cost $0.000534$. Training and evaluation methodology are identical to the baseline experiments.} We run two experiments to ascertain the performance benefits for contract-augmented agents. First, we train two contract-augmented agents and one A2C agent together. Figure \ref{fig:gifting-contracts}(A--C) shows that the two contract-augmented agents (Learners $0$ and $1$) are able to achieve a $2$-way draw and eliminate the agent without the ability to sign contracts. We then train three contract-augmented agents together. Figure \ref{fig:gifting-contracts}(D) demonstrates the reward dynamics that result as agents vie to make deals that will enable them to do better than a $3$-way draw. Figure \ref{fig:regression} shows that signing contracts has a significant correlation with accruing more chips in a game, thus demonstrating that contracting is advantageous. 

The benefits of contracting have an interesting economic interpretation. Without contracts, and the benefits of mutual trust they confer, there is no exchange of chips. In economic terms, the ``gross domestic product'' (GDP) is zero. Once agents are able to sign binding contracts, goods flow around the system; that is to say, economic productivity is stimulated. Of course, we should take these observations with a large grain of salt, for our model is no more than a toy. Nevertheless, this does hint that ideas from macro-economics may be a valuable source of inspiration for MARL algorithms that can coordinate and cooperate at scale. For instance, our model demonstrates the GDP benefits of forming corporations of a sufficiently large size $k > 1$ but also a sufficiently small size $k < 3$. 

\subsection{Contracts with temporal extent}

As it stands, our contract mechanism requires that contracts are fulfilled immediately, by invoking a legal actions mask on the agent's policy logits. On the other hand, many real-world contracts involve a promise to undertake some action during a specified future period; that is to say, contracts have temporal extent. Furthermore, directly masking an agent's action is a heavy-handed and invasive way to enforce that agents obey contractual commitments. By contrast, human contracts are typically enforced according to contract law \cite{10.2307/3657531}. Indeed, in the language of \cite{sep-legal-obligation} ``whatever else they do, all legal systems recognize, create, vary and enforce obligations''.

Legal systems stipulate that those who do not fulfil their contractual obligations within a given time period should be punished. Inspired by this, we generalize our contract mechanism. Contracts are offered and signed as before. However, once signed, agents have $b$ timesteps in which to take the promised action. Should an agent fail to do so, they are considered to have broken the contract, and receive a negative reward of $r_c$. Once a contract is either fulfilled or broken, agents are free to sign new contracts. 

\textbf{Trembling hand}. Learning to sign punishment-enforced contracts from scratch is tricky. This is because there are two ways to avoid punishment: either fulfil a contract, or simply refuse to sign any. Early on in training, when the acting policies are random, agents will learn that signing contracts is bad, for doing so leads to negative reward in the future, on average. By the time that the acting policies are sensibly distinguishing moves in different states of the game, the contract policies will have overfitted to never offering a contract. 

To learn successfully, agents must occasionally be forced to sign contracts with others. Now, as the acting policies become more competent, agents learn that contracts can be fulfilled, so do not always lead to negative reward. As agents increasingly fulfil their contractual obligations, they can learn to sign contracts with others. While such contracts are not binding in the sense of legal actions within the game, they are effectively binding since agents have learned to avoid the negative reward associated with breaking a contract. 

We operationalize this requirement by forcing agents to follow a \textit{trembling hand} policy for contract offers. Every timestep, the contract mechanism determines whether there are two agents not already participating in a contract. The mechanism chooses a random contract between the two, and applies a mask to the logits of the contract policy of each agent. This mask forces each agent to suggest the chosen contract with a given minimum probability $p_c$. We also roll out episodes without the trembling hand intervention, which are not used for learning, but plotted for evaluation purposes. 

\begin{figure}
    \centering
    \includegraphics[scale=0.38]{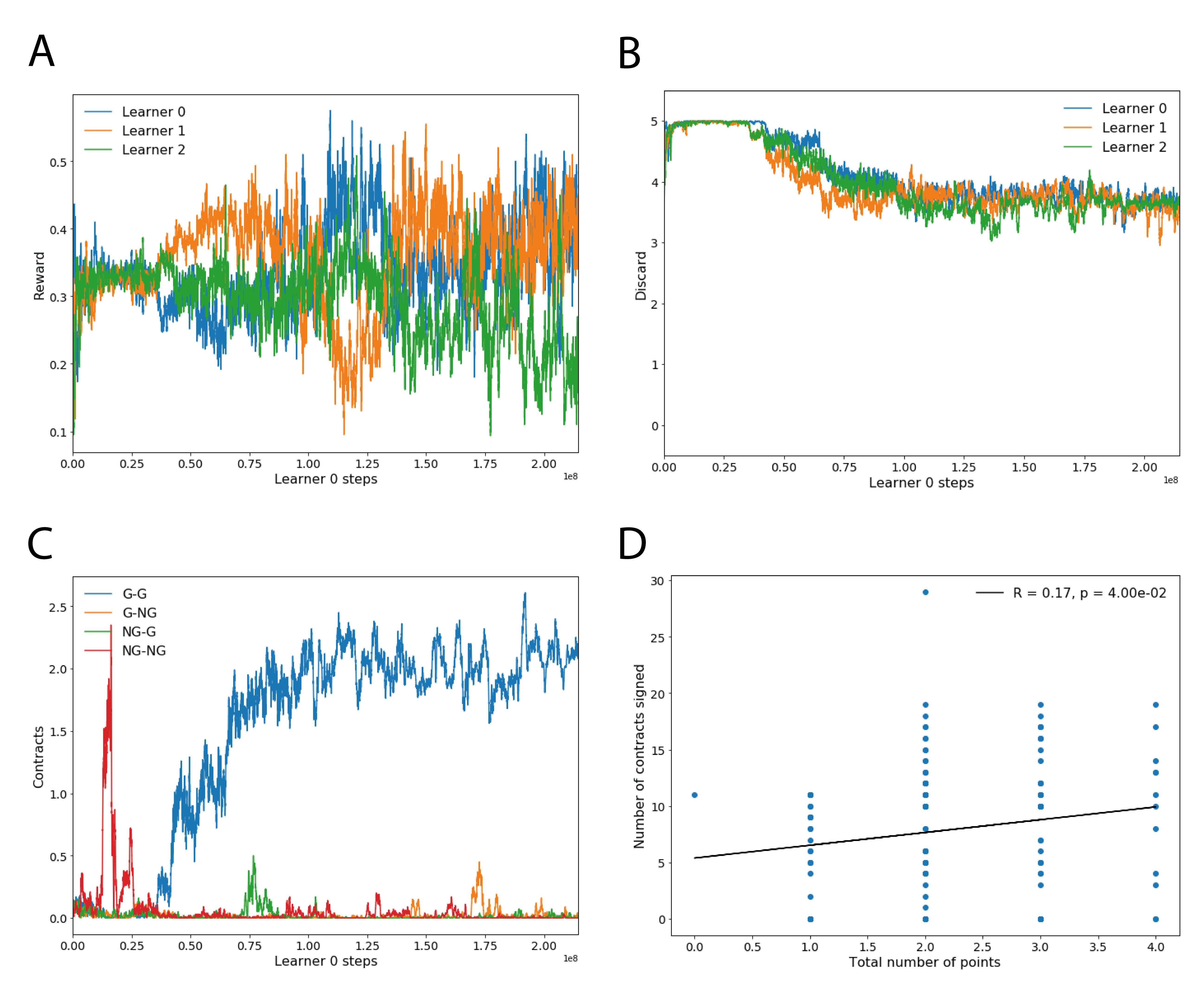}
    \caption{Punishment-enforced contracts lead to alliance formation in Gifting. (A) Initially agents achieve a $3$-way draw, receiving $\frac{1}{3}$ reward each. Later agents learn to form alliances in which two agents get reward $\frac{1}{2}$ and the other $0$. (B) After an initial period in which all agents discard, gifting behavior is learned. (C) Agents first learn to sign mutually non-gifting contracts (NG-NG), but later discover the benefit of mutually gifting contracts (G-G). (D) There is a significant positive correlation between number of contracts signed and points scored. Dependent variables are as follows. (A) Reward. (B) Discard. (C) Contracts signed. (D) Number of chips accrued vs. number of contracts signed.}
    \label{fig:punish-gift}
\end{figure}

\textbf{Results}.\footnote{The neural network is identical to the binding contracts experiment. For the punishment mechanism, we choose $b = 6$, $r_c = -1$ and $p_c = 0.5$. Training hyperparameters are as follows: learning rate $0.002738$, environment policy entropy cost $0.004006$, contract loss weight $3.371262$ and contract entropy cost $0.002278$. All other hyperparameters are unchanged from the previous experiments.} We ran one experiment to ascertain the ability of agents to learn to form gifting contracts. All agents are contract-aware and have trembling hand policies while learning. We display our results in Figure \ref{fig:punish-gift}. Initially agents learn to discard, and indeed back this up by signing contract for mutual discarding behavior. However, after some time agents discover the benefits of signing gifting contracts. Thereafter, the agents vie with each other to achieve two-player alliances for greater reward. Interestingly, the agents do not learn to gift as much as in the binding contracts case: compare Figure \ref{fig:punish-gift}(B) and Figure \ref{fig:gifting-contracts}(E). This is likely because contracts are not always perfectly adhered to, so there is a remnant of the fear and greed that drives discarding behavior. 

\section{Conclusion}

In this paper, we have made five key contributions. We (1) formalized one of the challenges of many-player zero-sum games by introducing the notion of an alliance dilemma. We (2) demonstrated that these are ubiquitous, and that (3) gradient-based and reinforcement learning fails to resolve them. We introduced (4) an approach for allowing MARL agents to form dynamic team, by augmenting the agents with a {\it binding contract} channel. These agents learn to use contracts as a mechanism for trusting a potential alliance partner. Finally, we (5) generalized our contract mechanism beyond the binding case, showing that agents learn to sign temporally-extended contracts enforced through punishment.

\textbf{Future work}. Our model suggests several avenues for further work. Most obviously, we might consider contracts in an environment with a larger state space, such a spatially extended gridworld. Here, agreeing on specific actions to take within an alliance is likely too granular. The alliance dilemma would emerge on the level of the meta-game defined by the policies of different players, defining a \textit{sequential alliance dilemma}. There are at least two promising approaches in such a setting. Firstly, we could incorporate a centralized agent to whose policy agents could defer, should they wish to enter an alliance. Secondly, we could make contracts about an abstraction over the state of the environment, rather than about atomic actions. Further, one might want to include multiple rounds of contract negotiation per time step of the environment, along the lines of \cite{williams2012}.

More generally, it would be fascinating to discover how a system of contracts might emerge and persist within multi-agent learning dynamics without directly imposing mechanisms for enforcement. Such a pursuit may eventually lead to a valuable feedback loop from AI to sociology and economics. Relatedly, we might ask how to scale contracts beyond the bilateral case, given the exponential explosion of possible alliances with the number of players in the game. Indeed, real-world institutions sign large numbers of contracts simultaneously, each of which may involve several partners. Finally, we note that our contract mechanism provides a simple grounded ``language'' for agent interaction. We hope to draw a stronger connection between this work and the emergent communication literature in the future.

\textbf{Outlook}. Many-player zero-sum games are a common feature of the natural world, from the economic competition of Adam Smith \cite{RePEc:hay:hetboo:smith1776} to Darwin's theory of evolution \cite{darwin1859} which can be viewed as a zero-sum game for energy \cite{van1980evolution}. In many-player zero-sum games a single agent cannot necessarily play a policy which is impervious to its opponent's behavior. Rather, to be successful, an algorithm must influence the joint strategy across many players. In particular, zero-sum multi-player games introduce the problem of dynamic team formation and breakup. This problem is remarkably deep, touching three strands of multi-agent research beyond two-player zero-sum games:

\begin{enumerate}
    \item Emergent teams must coordinate within themselves to effectively compete in the game, just as in team games like soccer.
    \item The process of team formation may itself be a social dilemma. Intuitively, players should form alliances to defeat others; however, membership of a alliance requires individuals to contribute to a wider good which is not completely aligned with their self-interest.
    \item Decisions must be made about which teams to join and leave, and how to shape the strategy of these teams. Here communication is vital, analogously to work on negotiation and contractual team formation.
\end{enumerate}

In Section \ref{sec:background} we identified three appealing features of zero-sum two-player games. Many-player zero-sum games inherit the appealing properties (1) and (2) from their two-player counterparts. However, the restrictive property (3) does not hold, since there is no general analogue of the minimax theorem beyond $2$-player games.\footnote{Although in restricted classes, a statement can be made \cite{Cai:2011:MTM:2133036.2133056}.} For example, the minimax-$Q$ algorithm \cite{Littman:1994:MGF:3091574.3091594} is not applicable, because it fails to account for dynamic teams. We have demonstrated that many-player zero-sum games capture intricate, important and interesting multi-agent dynamics, amenable to formal study with appropriate definitions. We look forward to future developments of AI in this fascinating arena of study.


\bibliographystyle{ACM-Reference-Format}  
\bibliography{sample-bibliography}  

\appendix

\section{Gradient-Based Learning}\label{appendix:fail}

We provide a simple mathematical argument that illuminates the failure of gradient-based learning in the Odd One Out and Matching games. In these games the equilibria for policy gradients are easy to characterize. For example in the Odd One Out game ($p=q=1$) if players use strategy $0$ with probabilities $x$, $y$ and $z$ respectively, then player $2$'s payoff is given by
\[ xyz/3 + (1-x)(1-y)(1-z)/3 +xy(1-z) + (1-x)(1-y)z  \, .\]

The policy gradient with respect to $z$ is $2(1-x-y)/3$, with fixed points when $x+y=1$ or on the boundary, where $z = 0$ or $1$. Therefore the fixed points are (a) $(0.5,0.5,0.5)$ and (b) $(1,0,z)$ and permutations. 

Note that (a) is unstable. With a softmax policy, (b) cannot be achieved exactly, but can be converged to. One player's policy converges to $1$, another's to 0, and the third's converges to either $0$ or $1$. The two players who have the same policy have thus failed to resolve their alliance dilemma.

\end{document}